\documentclass[aps,prx,twocolumn,superscriptaddress, preprintnumbers,amsmath,amssymb]{revtex4-1}
\usepackage{graphicx}
\usepackage{subfigure}
\usepackage{epsfig}
\usepackage{dcolumn}
\usepackage{bm}
\usepackage{ulem}
\usepackage{color}
\usepackage{verbatim}

\def\be{\begin{equation}}       \def\ee{\end{equation}}
\def\bea{\begin{eqnarray}}      \def\eea{\end{eqnarray}}
\def\ba{\begin{array}}
\def\ea{\end{array}}
\def\bnum{\begin{enumerate} }
\def\enum{\end{enumerate}}

\def\nn{\nonumber}

\def\=>{\Rightarrow}
\def\>{\rightarrow}

\def\eye2{Fathbb{I}}

\renewcommand{\>}{\rangle}

\begin{document}

\title{Fermion-induced quantum critical points in two-dimensional Dirac semimetals}

\author{Shao-Kai Jian}
\affiliation{Institute for Advanced Study, Tsinghua University, Beijing 100084, China}
\author{Hong Yao}
\email{yaohong@tsinghua.edu.cn}
\affiliation{Institute for Advanced Study, Tsinghua University, Beijing 100084, China}
\affiliation{State Key Laboratory of Low Dimensional Quantum Physics, Tsinghua University, Beijing 100084, China}
\affiliation{Collaborative Innovation Center of Quantum Matter, Beijing 100084, China}

\begin{abstract}
In this paper we investigate the nature of quantum phase transitions between two-dimensional Dirac semimetals and $Z_3$-ordered phases (e.g. Kekule valence-bond solid), where cubic terms of the order parameter are allowed in the quantum Landau-Ginzberg theory and the transitions are putatively first-order. From large-$N$ renormalization group (RG) analysis, we find that fermion-induced quantum critical points (FIQCPs) [Z.-X. Li {\it et al.}, Nature Communications {\bf8}, 314 (2017)] occur when $N$ (the number of flavors of four-component Dirac fermions) is larger than a critical value $N_c$. Remarkably, from the knowledge of spacetime supersymmetry, we obtain an {\it exact} lower bound for $N_c$, {\it i.e.}, $N_c>1/2$. (Here the ``1/2'' flavor of four-component Dirac fermions is equivalent to one flavor of four-component Majorana fermions). Moreover, we show that the emergence of two length scales is a typical phenomenon of FIQCPs and obtain two different critical exponents, i.e., $\nu$$\neq$$\nu'$, by large-$N$ RG calculations. We further give a brief discussion on possible experimental realizations of FIQCPs.
\end{abstract}
\date{\today}

\maketitle
		
\section{Introduction}

The Landau cubic-term criterion says that a phase transition must be first order if cubic terms are allowed in the Landau-Ginzberg free energy \cite{landau1958,lifshitz1942, binder1987,blote1979}. Transitions that violate the cubic-term criterion are rare; one classic example is the exactly solvable three-state Potts model in 1+1 dimension (1+1D) \cite{baxter1973, wu1982}. No higher dimensional example was identified until fermion-induced quantum critical points (FIQCPs) were introduced recently in Ref. \cite{li2015a}, where a putative first-order phase transition can be driven to be continuous by gapless Dirac fermions in 2+1D. In Ref. \cite{li2015a}, an SU($N$) fermionic model featuring a $Z_3$ transition from two-dimensional (2D) Dirac semimetals to Kekule valence-bond solids (VBSs), where cubic terms of the order parameters are allowed by symmetry, was conceived and studied using a large-scale sign-problem-free Majorana quantum Monte Carlo simulation (QMC) \cite{li2015b, li2016} and large-$N$ renormalization group (RG) calculations \cite{gross1974, colemanbook, rosenstein1993, moshe2003, herbut2006}. Both QMC simulations and RG analysis show the occurrence of a FIQCP when $N$, the number of flavors of four-component Dirac fermions, is larger than a critical value $N_c$. This FIQCP scenario was also confirmed later by $4\!-\!\epsilon$ and  functional RG calculations \cite{scherer2016, classen2017}.

\begin{figure}[t]
	\includegraphics[width=7.5cm]{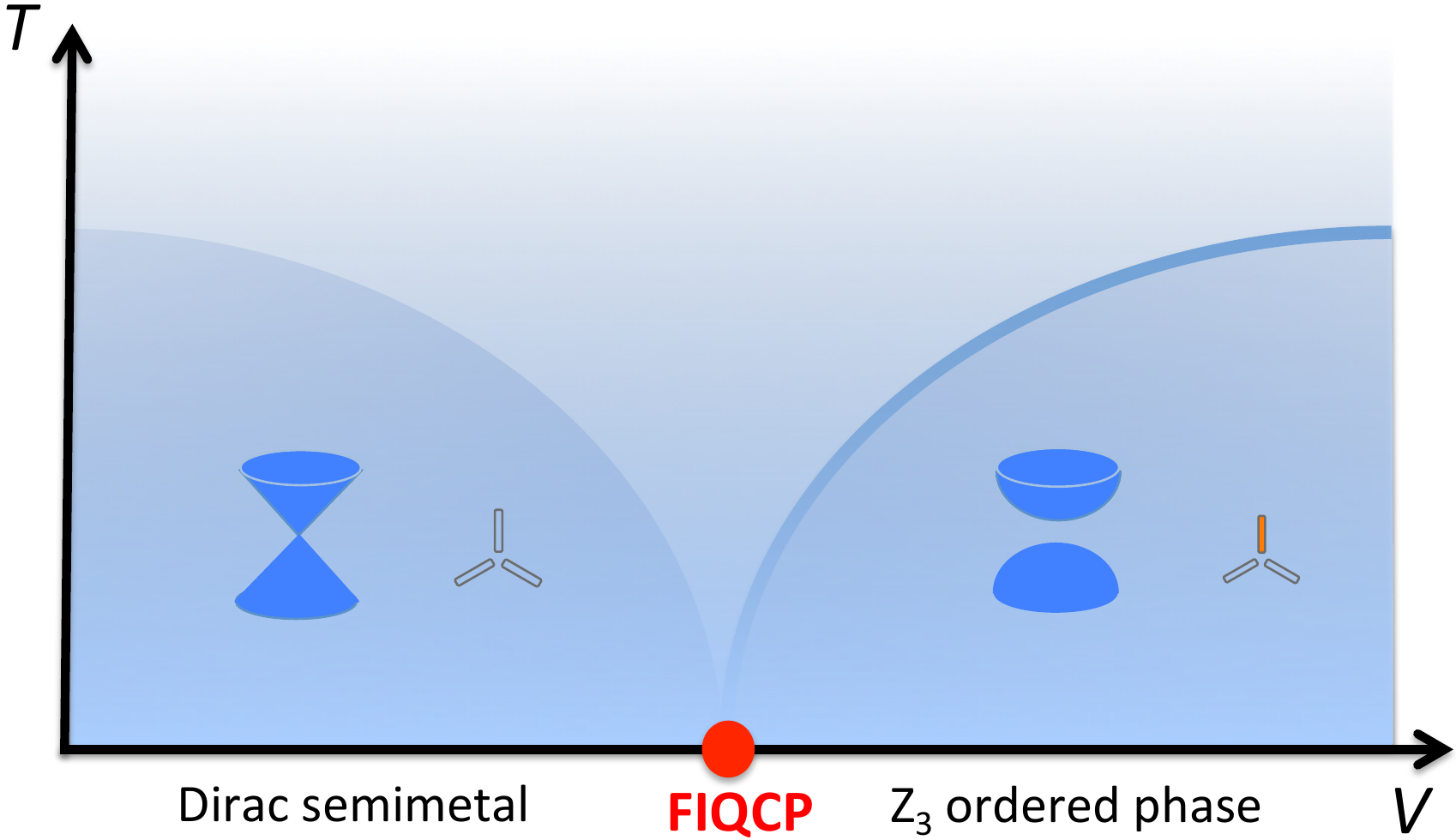}
	\caption{\label{fig1}A schematic phase diagram for a phase transition from a gapless Dirac semimetal to a $Z_3$ ordered phase. A fermion-induced quantum critical point (FIQCP) is presented in zero temperature.}
\end{figure}

It has been known that the number of flavors of Dirac fermions is often of crucial importance in determining the nature of low-energy physics and quantum phase transitions \cite{sachdevbook}. For example, the stability of algebraic spin liquids in 2+1D \cite{anderson1987, affleck1988, rantner2001, wen2002, wenbook, ryu2007, ran2007, yao2009, corboz2012, xiang2016} [equivalently, the stability of the deconfined phase of the compact quantum electrodynamics (QED) in 2+1D] depends essentially on the number of flavors of the emergent Dirac spinons \cite{polyakov1977, hermele2004, assaad2004, kleinert2005, xu2008, grover2014}; another related issue is the critical number of flavors of Dirac fermions for chiral symmetry breaking in non-compact QED \cite{grover2014, pisarski1984, appelquist1986, appelquist1988, nash1989, kondo1995, aitchison1997, herbut2002, fischer2005, bashir2008, strouthos2009, gusynin2016}.  For the quantum phase transitions between the 2D Dirac semimetals and a broken $Z_3$ symmetry phase with massive Dirac fermions, which are called $Z_3$ Gross-Neveu-Yukawa transitions, the large-$N$ one-loop RG calculations \cite{li2015a} give rise to $N_c\!=\!\frac12$ such that only when $N\!>\!N_c$ can a FIQCP occur. A schematic quantum phase diagram for the occurrence of such a FIQCP is shown in Fig. \ref{fig1}. Large-$N$ RG calculations show that FIQCP cannot occur at $Z_3$ Gross-Neveu-Yukawa transitions in 3+1D for any finite $N$.

Nonetheless, obtaining non-perturbative and even exact results about $N_c$, the critical number of flavors of Dirac fermions for the occurrence of FIQCPs at $Z_3$ Gross-Neveu-Yukawa transitions in 2+1D, is desired. In this paper, with the help of spacetime supersymmetry (SUSY) \cite{wessbook,weinbergbook, seiberg1994, aharony1997, strassler2003}, we can show {\it exactly} that for $N=\frac12$ such a $Z_3$ Gross-Neveu-Yukawa transition must be first-order. Consequently, we obtain an exact lower bound for $N_c$, namely, $N_c\!>\!\frac12$.

Another interesting aspect of FIQCPs is the emergence of two length scales \cite{nelson1976, miyashita1997, oshikawa2000, carmona2000, okubo2015, leonard2015, shao2016} due to the presence of dangerously irrelevant operators at the transition point, which is a typical phenomenon of Landau-forbidden phase transitions. For instance, two length scales are found at deconfined quantum critical points (DQCPs) \cite{senthil2004a, senthil2004b, lou2007, ghaemi2010, nahum2015, shao2016, kaul2013}. FIQCPs, as a scenario of Landau-forbidden transitions, also exhibit the emergence of two length scales owing to the dangerously irrelevant cubic terms at the critical points. A consequence of two emergent length scales is that critical exponents can be different on the two sides of the transition. We obtain these different critical exponents with one-loop RG calculations.

This paper is organized as follows. In Sec. II, the low-energy effective theory describing the $Z_3$ phase transitions is constructed by symmetry and scaling considerations. In Sec. III, a large-$N$ RG analysis is carried out for the $Z_3$ Gross-Neveu-Yukawa transitions in $d$+1 dimensions, where $N$ is the number of four-component Dirac fermions and $2\le d \le 3$. Critical spatial dimensions for the occurrence of $Z_3$ Gross-Neveu-Yukawa transitions are discussed. One-loop RG analysis shows that FIQCPs can occur for $N>1/2$ when $d=2$ but cannot occur for any $N$ when $d=3$. In Sec. IV, we construct field theory with a single Dirac cone (two Majorana cones), i.e., $N=\frac12$, interacting with the $U(1)$ order parameter in 2+1D \cite{roy2013}, where the FIQCP is assumed to occur by first ignoring the cubic anisotropy. The fixed point theory has emergent SUSY in 2+1D \cite{ashvin2014, ponte2014, jian2015a, jian2016b, joseph2016, grover2016, li2017a}, corresponding to the SUSY Wess-Zumino model after a particle-hole transformation. However, cubic perturbations at such a SUSY fixed point turn out to be relevant, rendering the phase transition discontinuous and ruining the assumed FIQCP. Consequently, this provides an exact lower bound, $N_c\!>\!\frac12$, for the occurrence of FIQCP at $Z_3$ Gross-Neveu-Yukawa transitions in 2+1D. In Sec. V, we investigate two emergent length scales at the FIQCP. We obtain different critical exponents on the two sides of the transition. In the end, we conclude these theoretical investigations and give a brief discussion of possible experimental realizations of FIQCPs in Sec. VI.

\section{The effective theory for $Z_3$ Gross-Neveu-Yukawa transitions in Dirac semimetals}

The noninteracting Dirac fermion in $d$ dimensions ($d$ is the spatial dimension) is well known and given by
\bea\label{dirac}
	\mathcal{L}_\psi &=& \psi^\dag ( \partial_\tau -i v \gamma^\mu \partial_\mu)\psi,
\eea
where summation over $\mu=1,...,d$ is assumed implicitly and the gamma matrices are $2^{[\frac{d}{2}+1]}$-dimensional, where $[x]$ denotes the integer part of $x$, and satisfy $\gamma^{\mu\dag}=\gamma^\mu$, $\{ \gamma^\mu,\gamma^\nu \}=2\delta^{\mu\nu}$. Here $v$ is the Fermi velocity, and $\tau$ is imaginary time. The Dirac fermions presented here have rotational symmetry, i.e., the Fermi velocity is assumed to be isotropic, which is an emergent symmetry in the infrared \cite{vafek2002,isobe2012,roy2016}. A concrete 4$\times$4 matrix representation of $\gamma$ matrices in $d\!=\!2$ can be given by $\gamma^1=\sigma^x \tau^z,\gamma^2=\sigma^y$, where $\sigma$ and $\tau$ are Pauli matrices. For spinless fermions in a 2D honeycomb lattice with nearest neighbor (NN) hopping, the Pauli matrices $\sigma$ ($\tau$) represent sublattice (valley) indices.

To gap out Dirac fermions by the $Z_3$ order (two-dimensional representation), $d\!+\!2$ anti-commuting $\gamma$ matrices are necessary: $d$ gamma matrices represent the relevant degrees of freedom of the $d$-dimensional Dirac fermions shown above, while the other two $\gamma$ matrices, $\gamma^{d+1}$ and $\gamma^{d+2}$, fulfill two mass terms related to the $Z_3$ order parameters \cite{ryu2009,hou2007, ghaemi2010}. In this representation, the $Z_3$ symmetry is generated by $T\!=\! \exp[\frac{\pi}{3} \gamma^{d\!+\!1} \gamma^{d\!+\!2}]$. Thus, the lowest-order coupling term between Dirac fermions and the order parameters is
\bea
	\mathcal{L}_{\psi\phi} &=& g(\phi \psi^\dag \gamma^+ \psi +\phi^\ast \psi^\dag \gamma^- \psi ) ,
\eea
where $\phi$ is a complex order parameter, and $\gamma^\pm \!=\!\frac{1}{2}(\gamma^{\!d+\!1}\! \pm \! i\gamma^{d\!+\!2})$, and $g$ is a real constant describing the strength of fermion-boson coupling. The fermion-boson coupling has $O(2)$ symmetry larger than $Z_3$. For the transition between the Dirac semimetal and Kekule VBS on the honeycomb lattice, two $\gamma$ matrices are $\gamma^3 \!=\! \sigma^x \tau^x,\gamma^4\!=\!\sigma^x \tau^y$. The order parameter is given by the Kekule valence-bond density with $2\vec K$ momentum, where $\vec K$ is the valley momentum and the $Z_3$ symmetry corresponds to the translation symmetry in the honeycomb lattice.

For the Landau-Ginzberg (LG) theory of the order parameter, the Lagrangian dictated by symmetries reads
\bea
	\mathcal{L}_\phi &=& |\partial_\tau \phi|^2 \!+\! c^2 |\nabla \phi|^2 + r|\phi|^2 +b(\phi^3\!+\!\phi^{\ast3}) + u|\phi|^4, \label{boson1}
\eea
where $r$ is the effective tuning parameter for phase transition, $b$ is a real constant representing the strength of cubic terms and $u$ is assumed to be positive. Again, the isotropy of boson velocity $c$ is a low-energy emergent phenomenon. Note that the cubic term is allowed by symmetries and lowers the $O(2)$ symmetry of LG theory to $Z_3$ and putatively renders a first-order phase transition according to the Landau cubic-term criterion. An effective particle-hole symmetry ($\phi\! \rightarrow\! \phi^*$, e.g., reflection symmetry in the Kekule transition) is also assumed to exclude the first-order time derivative of the order parameters.

\section{Large-$N$ \label{rg}Renormalization group analysis for $Z_3$ Gross-Neveu-Yukawa transitions in Dirac semimetals}

\subsection{RG equations in $d+1$ dimensions}
\begin{figure}[b]
	\subfigure[]{\label{kb1}
		\includegraphics[width=2.6cm]{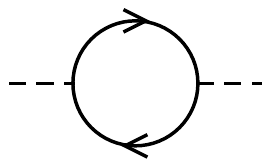}}
	\subfigure[]{\label{kb2}
		\includegraphics[width=2.7cm]{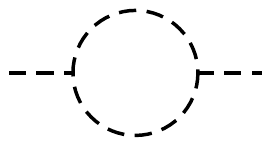}}
	\subfigure[]{\label{kf}
		\includegraphics[width=2.7cm]{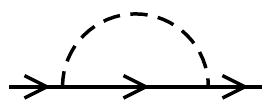}}
	\subfigure[]{\label{3b}
		\includegraphics[width=2.6cm]{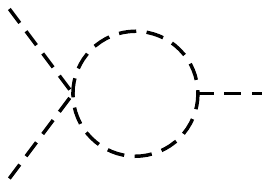}}
	\subfigure[]{\label{4b1}
		\includegraphics[width=2.6cm]{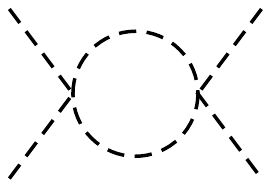}}
	\subfigure[]{\label{4b2}
		\includegraphics[width=2.5cm]{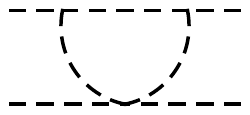}}
	\subfigure[]{\label{4b3}
		\includegraphics[width=2.2cm]{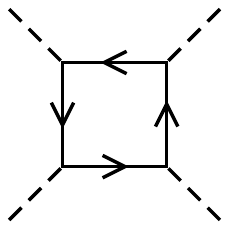}}
	\subfigure[]{\label{4b4}
		\includegraphics[width=2.2cm]{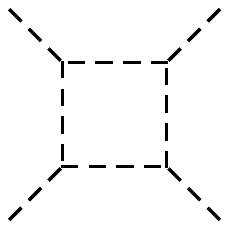}}
	\caption{\label{fig2} One-loop Feynman diagrams. The arrowed solid line indicates the fermion propagator and the dashed line indicates the boson propagator.}
\end{figure}

The ultraviolet degrees of freedom (fast modes) constrained in the momentum shell $\Lambda e^{-l} \!<\!p\!<\!\Lambda$, where $l\!>\!0$ is a flow parameter and $\Lambda$ is a cutoff, are integrated out to generate the RG equations. In the following calculations, we assume the four-component Dirac fermions have $N$ flavors and use 4$\times$4 $\gamma$ matrices for simplicity since we are only interested in only $d=2,3$. Various trace properties for $\gamma$ matrices are given in Appendix \ref{trace}. Non-vanishing one-loop Feynman diagrams are shown in Fig. \ref{fig2}. The critical $r_c$ is renormalized to be $-\frac{8}{N}$ and will not affect the fixed point properties at the lowest level(see Appendix \ref{mass_rg} for details). So we set $r\!=\!0$ in the following calculations for simplicity.

The renormalizations of boson and fermion self-energy come from Figs. \ref{kb1}, \ref{kb2}, and \ref{kf}.  After evaluating the Feynman diagrams in Figs. \ref{kb1} and \ref{kb2}, the boson self-energy reads
\bea
\Pi(p)&=& g^2 \frac{N \pi }{2v^3} K_d (\omega^2+ v^2 p^2) \nn  +b^2 \frac{9\pi}{4c^5} K_d\Lambda^{d-5}l (\omega^2-\frac{1}{3}c^2 p^2 ),\\
\eea
where $K_d \!\equiv\! \frac{A_{d}}{(2\pi)^{d+1}}$, $A_{d}$ is the area of the unit $d$-sphere, and $p^2\!\equiv\!\sum_{\mu=1}^d p_\mu^2$. The Feynman diagram in Fig. \ref{kf} gives rise to fermion self-energy,
\bea
\Sigma(p)= \frac{g^2 \pi  K_d\Lambda^{d-3}l}{c(c+v)^2} (-i \omega+\frac{2c+v}{3v}\gamma^\mu p_\mu).
\eea
From the self-energy parts, the RG equations for boson and fermion velocities are given by
\bea
	\frac{dc}{dl} &=& -\tilde{g}^2 \frac{N\pi(c+v)}{4 c v^3} (c-v)-\tilde{b}^2 \frac{3\pi}{2 c^4}, \\
	\frac{dv}{dl} &=& -\tilde{g}^2 \frac{2 \pi(v-c)}{3cv(c+v)^2},
\eea
where dimensionless coupling constants are defined as $\tilde{g}^2\!\equiv\! K_d \Lambda^{d-3} g^2,\tilde{b}^2\!\equiv\! K_d \Lambda^{d-5} b^2, \tilde{u} \!\equiv\! K_d \Lambda^{d-3} u$. The anomalous dimensions for both boson and fermion fields are $ \eta_\phi \!=\!\frac{N\pi \tilde{g}^2}{4v^3}\!+\! \frac{9\pi\tilde{b}^2}{8c^5} , \eta_\psi \!=\! \frac{\pi\tilde{g}^2}{2c(c+v)^2}$. For a continuous phase transition, one demands $\tilde{b}^2\!=\!0$ at the fixed point. In such case, the velocities of fermions and bosons will flow to the same value, as seen in the following RG equations,
\bea
	\frac{d(c-v)}{dl}\!=\!-\tilde{g}^2 \Big[\frac{N\pi(c+v)}{4 c v^3} \!+\! \frac{2 \pi}{3cv(c+v)^2} \Big] (c-v).
\eea

The rest of the Feynman diagrams generate the RG equations for coupling constants. From the Feynman diagram in Fig. \ref{3b}, the renormalization for three-boson vertex is
\bea
\Gamma_{\phi^3} = \Gamma_{\phi^{\ast3}}= -b u  \frac{3\pi}{c^3} K_d \Lambda^{d-3}l.
\eea
By evaluating Figs. \ref{4b1}, \ref{4b2}, \ref{4b3}, and \ref{4b4}, we get
\bea
\Gamma_{|\phi|^4} &=&-u^2 \frac{5\pi}{c^3} K_d \Lambda^{d-3} l +  u b^2\frac{54\pi}{c^5} K_d \Lambda^{d-5} l \nn\\
	&&+ g^4 \frac{N \pi}{2v^3} K_d\Lambda^{d-3} l   -b^4\frac{405 \pi}{4 c^7} K_d \Lambda^{d-7} l .
\eea
Combining all vertices, we arrive at a set of RG equations for coupling constants,
\bea
 \frac{d \tilde{g}^2}{dl} &=& (3-d) \tilde{g}^2- \frac{9\pi}{4c^5}\tilde{b}^2 \tilde{g}^2-\Big[\frac{N\pi}{2v^3}+ \frac{2\pi}{c(c+v)^2}\Big] \tilde{g}^4, \nn\\
 \\
 \frac{d \tilde{b}^2}{dl} &=& (5-d)\tilde{b}^2 -\frac{3N\pi}{2v^3} \tilde{g}^2 \tilde{b}^2-\frac{6\pi}{c^3} \tilde{b}^2 \tilde{u}- \frac{27\pi}{4c^5} \tilde{b}^4, \\
 \frac{d \tilde{u}}{dl}&=&(3-d)\tilde{u}-\frac{N\pi}{v^3} \tilde{g}^2 \tilde{u} +\frac{N \pi}{2v^3}\tilde{g}^4 \nn\\
&&~~~~~~~~~~~~~~~~ - \frac{5\pi}{c^3}\tilde{u}^2 +\frac{99\pi}{2 c^5}\tilde{u} \tilde{b}^2  -\frac{405\pi}{4c^7} \tilde{b}^4.
\eea

\subsection{Prediction of FIQCP for $N>1/2$ and $d=2$}

\begin{figure}
	\includegraphics[width=7.5cm]{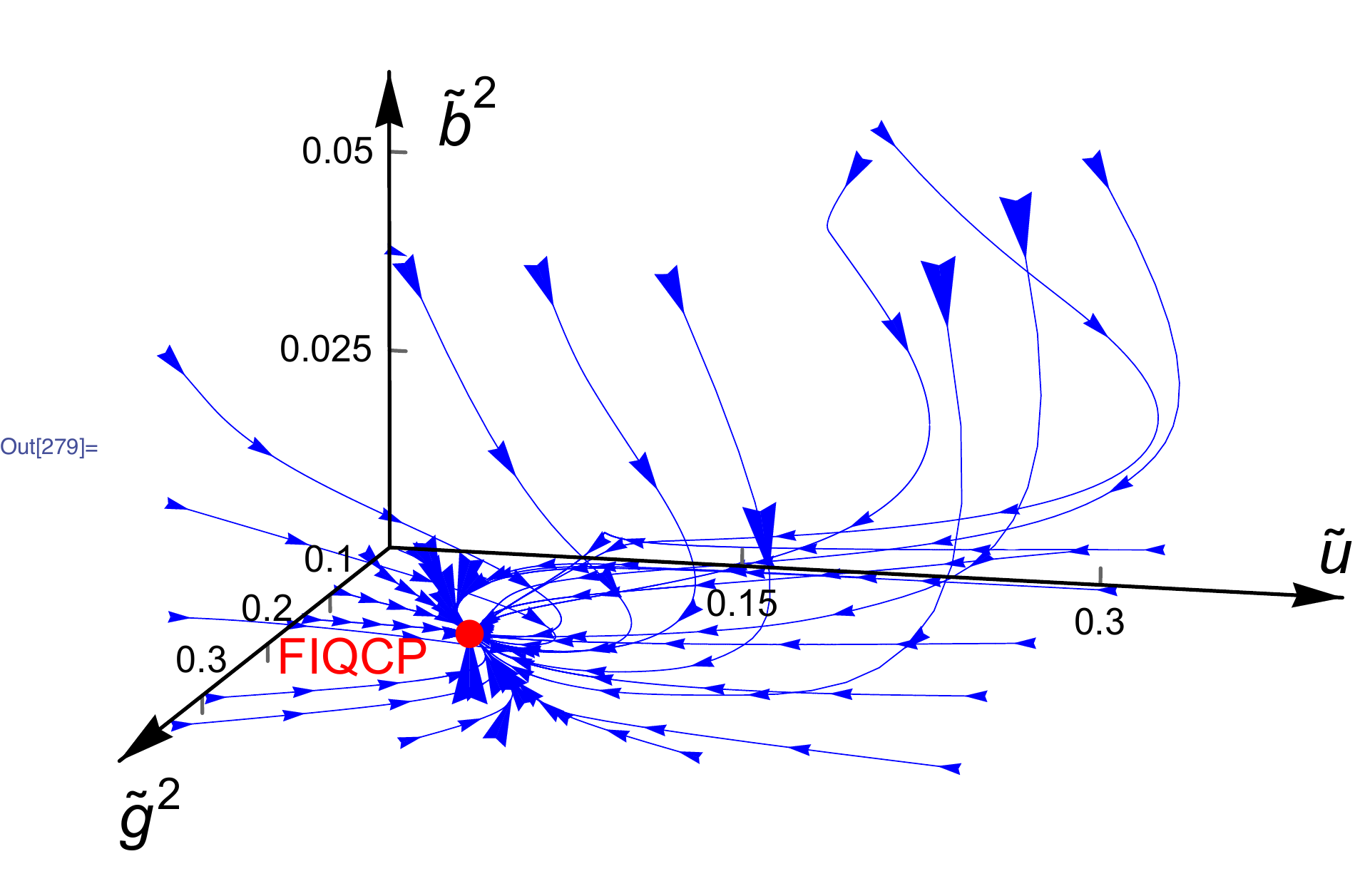}
	\caption{\label{fig3} The flow diagram in three dimensions generated by the RG equations for $N\!=\!3$. The blue arrows indicate the flow of coupling constants when energy is lowered. The red point located in the $\tilde{g}^{2}$-$\tilde{u}$ plane is a stable fixed point corresponding to the FIQCP.}
\end{figure}

When $d=2$ (two spatial dimensions), the RG flow equations above have a nontrivial fixed point ($\tilde{g}^{*2}$,$\tilde{u}^*$,$\tilde{b}^{*2}$) = ($\frac{2}{(\!N\!+\!1\!)\pi}$,$\frac{\sqrt{N^2+38N+1}\!-\! N\!+\!1}{10(N+1)\pi}$,0), where we have set $v^*\!=\!c^*\!=\!1$ for simplicity since the boson velocity and the fermion velocity flow to the same value at this fixed point, as discussed above. In the large-$N$ limit, the fixed point approaches ($\frac{2}{N\pi}$,$\frac{2}{N\pi}$,0), well controlled by $\frac{1}{N}$. A crucial feature is the emergent $O(2)$ symmetry at this fixed point; in other words, the cubic terms of the order parameter vanish, which is essential for the occurrence of a continuous transition point. By linearizing RG equations near this fixed point, we get the eigenvalues of the scaling matrix, which are  ($-1$,$-\frac{\sqrt{N^2+38N+1}}{N+1}$,$\frac{3(N+4-\sqrt{N^2+38N+1})}{5(N+1)}$). The first two are always negative while the last eigenvalue is positive for $N\!<\!\frac12$ and negative for $N\!>\!\frac12$, indicating the existence of a critical $N_c$ such that if the number of Dirac fermion flavors $N$ is larger than $N_c$, the fixed point is stable. The RG flow diagrams in the coupling-constant space spanned by ($\tilde{g}^{*2}$,$\tilde{u}^*$,$\tilde{b}^{*2}$) as shown in Fig. \ref{fig3} for $N\!=\!3$ indicate a stable fixed point consistent with our analysis.

A stable fixed point at critical surface $r\!=\!r_c(\tilde{g}^2,\tilde{u},\tilde{b}^2)$ corresponds to a genuine critical point. Since we have identified a stable fixed point with zero $\tilde{b}^{*2}$, i.e., no cubic terms present in this fixed point, it clearly corresponds to a FIQCP. The presence of gapless Dirac fermions at the transition point renders the cubic term irrelevant and results in a FIQCP violating the Landau criterion. The critical exponent $\eta$ at this critical point is given by $\eta \!\equiv \!2\eta_\phi\!=\!\frac{N}{N+1}$. Due to the existence of Dirac fermions at the transition, the dynamical exponent is $z\!=\!1$. And the fermion anomalous dimension is $\eta_\psi\!=\!\frac{1}{4(N+1)}$. The non-vanishing fermion anomalous dimension indicates the breakdown of the quasiparticle picture, and the critical theory is strongly interacting.

\begin{figure}
	\subfigure[]{
		\includegraphics[width=3.2cm]{boson_self2}}~~~~~~~~
	\subfigure[]{
		\includegraphics[width=2.8cm]{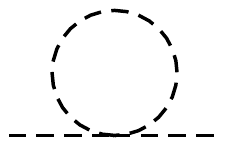}}
	\caption{\label{fig4} One-loop Feynman diagrams that renormalize the boson mass.}
\end{figure}

To explore other critical behavior, we also calculate the renormalization of boson mass (the relevant direction). Relevant Feynman diagrams are shown in Fig. \ref{fig4}. A straight forward calculation gives the boson self-energy that renormalizes the mass term, i.e.,
\bea
	\Gamma_{|\phi|^2} &=&\left(- \frac{9\pi\tilde{b}^2}{(1+\tilde{r})^{3/2}} + \frac{4\pi\tilde{u}}{\sqrt{1+\tilde{r}}} \right) \Lambda^2 l ,
\eea
where $\tilde{r}\equiv \Lambda^{-2} r$ is a dimensionless constant. The scaling dimension for $r$ at the critical point is $y_r\!=\!2-2\eta_\phi+ \frac{27\pi}{2} \tilde{b}^{\ast2}-2\pi\tilde{u}^\ast\!=\! 2\!-\frac{\sqrt{N^2+38N+1}+4N+1}{5(N+1)}$ and the critical exponent is given by $\nu^{-1}=y_r$. Other critical exponents can be determined by hyperscaling relations.

\subsection{No FIQCP for any $N$ and $d=3$}

\begin{figure}
	\includegraphics[width=6.5cm]{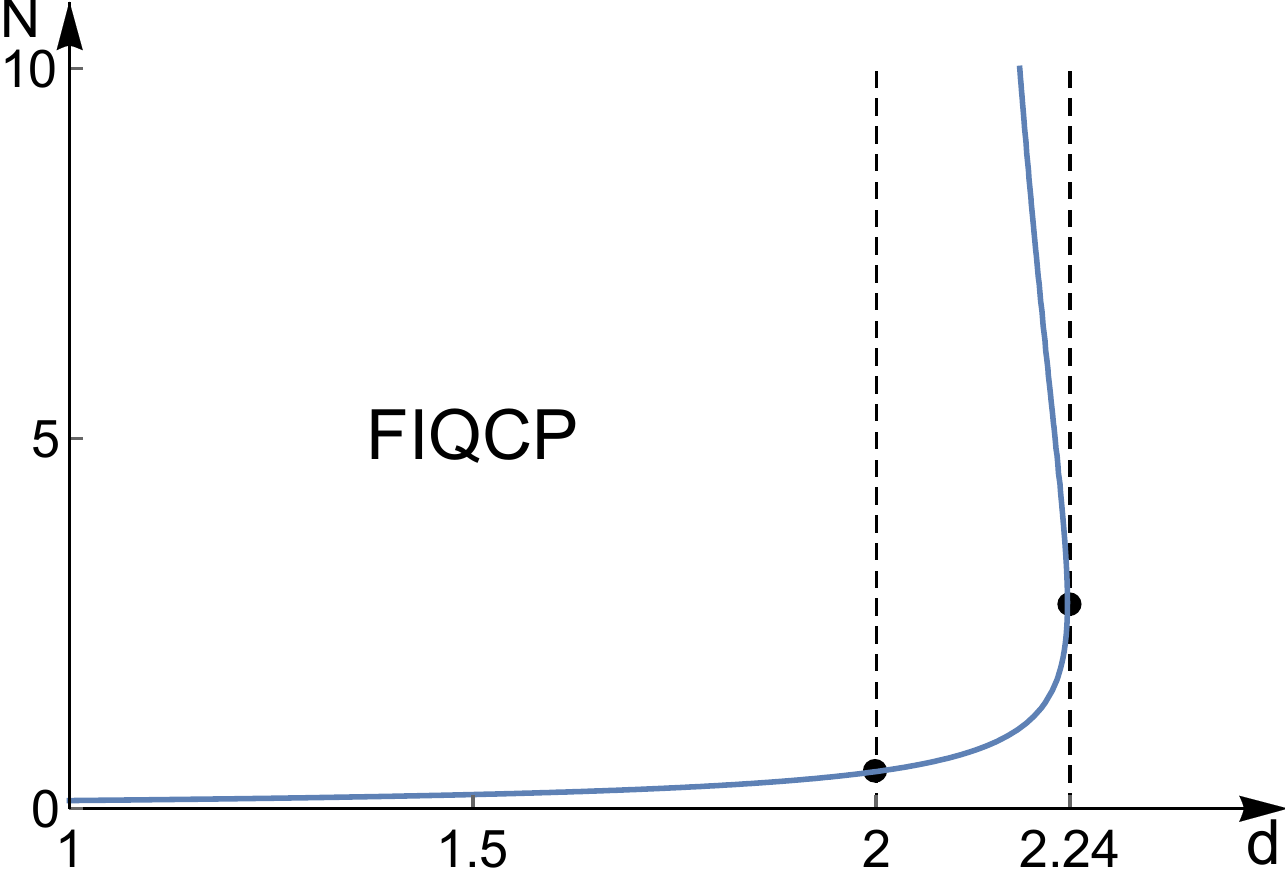}
	\caption{\label{fig5} Critical number of Dirac fermion flavors at $d$ dimensions. The curve corresponds to $N_l(d)$ and $N_u(d)$ and separates the region where FIQCP occurs. $N_l(2)\!=\!\frac12$ and $d_c\approx 2.24$ are shown explicitly on the curve.}
\end{figure}

Having explored the properties of FIQCP in 2+1D, we discuss the RG analysis in generic dimensions. Mean-field theory indicates the upper critical dimension for the existence of such a FIQCP is $d_{\text{MF}}\!\equiv\!2$ \cite{li2015a}. Our RG result is consistent with the mean field calculation since the mean field result corresponds to $N\!\rightarrow\! \infty$. Inclusion of quantum fluctuations for finite $N$ beyond mean-field theory will lead to a correction to the upper critical dimension, as we discuss below.

The solution of the RG equations in $d\!<\!3$ dimensions with both $\tilde{g}^{*2}\!>\!0$ and $\tilde{u}^*\!>\!0$ reads
\bea
&&\tilde{g}^{*2}=\frac{2(3\!-\!d)}{(N\!+\!1)\pi},\\
&&\tilde{b}^{*2}=0, \\
&& \tilde{u}^*\!=\!\frac{(3\!-\!d)(\sqrt{N^2+38N+1}\! -\!N \!+\!1)}{10(N+1)\pi}.
\eea
To determine whether the solution corresponds to a FIQCP, we calculate the eigenvalues of the scaling matrix by linearizing the RG equations near the fixed point. They are given by ($d\!-\!3$,$\frac{(d\!-\!3)\sqrt{N^2+38N+1}}{N+1}$,$\frac{16-11N+d(7N-2)-3(d\!-\!3)\sqrt{N^2+38N+1}}{5(N+1)}$). Although the first two scaling fields are always negative when $d\!<\!3$, the third scaling field could be positive depending on the number of flavors $N$ and the dimensionality $d$. We find a lower bound of the number of Dirac fermions for the occurence of the FIQCP in $d$ dimensions, i.e.,
\bea
N_l(d)\!=\!\frac{37d^2\!-\!232d\!+\!343\!-\!9\sqrt{(d\!-\!3)(17d^2\!-\!110d\!+\!161)}}{8d^2-20d+8}, \nn \\
\eea
and also an upper bound when $d>d_{\text{MF}}$,
\bea
N_u(d)\!=\!\frac{37d^2\!-\!232d\!+\!343\!+\!9\sqrt{(d\!-\!3)(17d^2\!-\!110d\!+\!161)}}{8d^2-20d+8}. \nn \\
\eea
The existence of an upper bound in $d>d_{\text{MF}}$ is expected from the mean-field results since the RG calculation should reproduce the mean-field result when $N\rightarrow \infty$. These two bounds merge at $d_c\!=\!\frac{55\!-\!12\sqrt{2}}{17}\!\approx\! 2.24$. The region for the FIQCP to occur is shown in Fig. \ref{fig5}. Thus, for $d=3$, Dirac fermions cannot induce a FIQCP for any $N$.  Nonetheless, it was shown that a FIQCP can occur at $Z_3$ nodal-nematic transitions in 3+1D double-Weyl systems \cite{jian2016a}.

\section{An exact result: No FIQCP for $N=1/2$ and $d=2$}

We now prove an exact result for the absence of FIQCP for $N=1/2$ and $d=2$ by employing the results of spacetime supersymmetry. The low-energy physics of Majorana fermions on a honeycomb lattice with next-neighbor hopping is equivalent to a single two-component Dirac fermion or half flavor of four-component Dirac fermions ($N\!=\!\frac12$). The noninteracting Hamiltonian reads
\bea
	H= \sum_{\langle xx' \rangle}  (i t_{xx'}\gamma_x \gamma_{x'}+ h.c.),
\eea
where $\langle xx'\rangle$ represents hopping between NN sites, the NN hopping amplitude $t_{xx'}=t$, and $\gamma_x$ is a Majorana operator at site $x$ satisfying $\gamma^\dag\!=\!\gamma$, and $\{ \gamma_x,\gamma_{x'} \}\!=\!2\delta_{xx'}$.  Using Fourier transformation, $\gamma_x \!=\!\frac{1}{\sqrt{2M}}\sum_{\vec k} \gamma_{\vec k} e^{i \vec k\cdot \vec x}$, where $M$ is the site number, we obtain
\bea
	H=\frac{t}{2} \sum_{\vec k}  [i \gamma_{A,-\vec k} \gamma_{B,\vec k} (1+e^{-i\vec k\cdot \vec e_1}+e^{-i\vec k\cdot \vec e_2})+ h.c.],~~~
\eea
where $\vec e_1\!=\!(\frac12,\frac{\sqrt{3}}{2})$, $\vec e_2\!=\!(-\frac12,\frac{\sqrt{3}}{2})$ are lattice vectors (lattice constant is set to be 1 for simplicity), and $A,B$ are sublattice indices. Diagonalizing the Hamiltonian, we get two Majorana cones $\psi_\pm$ located at the $\pm \vec K\!=\!(\pm \frac{4\pi}{3},0)$ points. The low-energy noninteracting Hamiltonian is given by
\bea
	\mathcal{H}= \psi^\dag v(i \sigma^x \partial_y-i \sigma^y \tau^z \partial_x ) \psi,
\eea
where $\psi\!\equiv\!(\psi_+,\psi_-)^T$, $v\!\equiv\! \frac{\sqrt{3}t}{4}$ is the Fermi velocity, and $\sigma$ and $\tau$ are Pauli matrices with sublattice indices $A, B$ and valley indices, respectively.

Note that only half of the Majorana fermions in $k$ space are independent, since $\gamma_{\vec k}^\dag \!=\! \gamma_{-\vec k}$. In other words, two Majorana cones are related through $\psi_{-}\!=\!\psi_+^\dag$ such that the two Majorana cones can be combined into a single Dirac cone. Then, we consider the Kekule VBS order, for which the order parameter is given by $\phi \!\propto\! \psi^\dag_- \sigma^y \psi_+\!\equiv\! \psi_+^T \sigma^y \psi_+$, which is equivalent to intravalley ``pairing'' of a single Dirac cone. Note that there is not $U(1)$ corresponding to particle number conservation symmetry in such a theory because of the real nature of Majorana fermions. Instead of being a pairing order parameter, $\phi$ actually breaks the translational symmetry of the honeycomb lattice since it carries a finite momentum $2\vec K$ and it is $Z_3$ symmetry breaking.

We first assume that the phase transition is continuous by setting $b\!=\!0$ to get a fixed-point theory and then determine the fate of cubic terms at this fixed point. Setting $b\!=\!0$, the effective theory after a particle-hole transformation at the assumed critical point is
\bea
	\mathcal{L} &=& \psi^\dag [\partial_\tau \!+v(\!i \sigma^x \partial_y\!-\!i \sigma^y \partial_x)] \psi+ |\partial_\tau \phi|^2 +c^2 \sum_i |\partial_i \phi|^2 \nn\\
	&&~~+ \frac g2 (\phi \psi^T \sigma^y \psi + h.c.)+ u|\phi|^4 ,
\eea
where $\psi$ represents a single two-component Dirac fermion. The factor $\frac12$ in the coupling constant $g$ comes from the fact that we rescale $\psi$ to $\frac{\psi}{\sqrt{2}}$ when combining two Majorana cones into one Dirac cone. Setting $N=\frac12$ in the RG flow equations, we obtain the following fixed point: $c^\ast\!=\!v^\ast, (\frac{g^\ast}{2})^2\!=\!u^\ast$. Such a critical theory has emergent SUSY and corresponds to the $\mathcal{N}\!=\!2$ Wess-Zumino model \cite{wessbook, weinbergbook, ashvin2014,ponte2014}.

We showed earlier that when $N\!=\!\frac12$, the cubic terms are marginal at the one-loop level, and could be marginally relevant or irrelevant if higher-order quantum fluctuations are included. Interestingly, owing to the $U(1)_{\mathcal{R}}$ symmetry in the $\mathcal{N}=2$ Wess-Zumino model, the scaling dimension of chiral multiplets is known {\it exactly} \cite{seiberg1994,aharony1997}, and is equal to the $\mathcal{R}$-charge. Thus, the scaling dimension of $\phi^3$ at the assumed fixed point is known exactly: $\Delta_{\phi^3}\!=\!2$, which is much smaller than 3. As a result, the cubic term is strongly relevant, and the assumed fixed point is unstable due to the relevance of cubic perturbations. Namely, the $N=\frac12$ flavor of four-component Dirac fermions cannot induce a FIQCP at the $Z_3$ Gross-Neveu-Yukawa transition. In other words, the critical number of flavors of Dirac fermions satisfies $N_c>\frac12$. This inequality is {\it exact}!

\begin{figure}[t]
	\includegraphics[width=8.5cm]{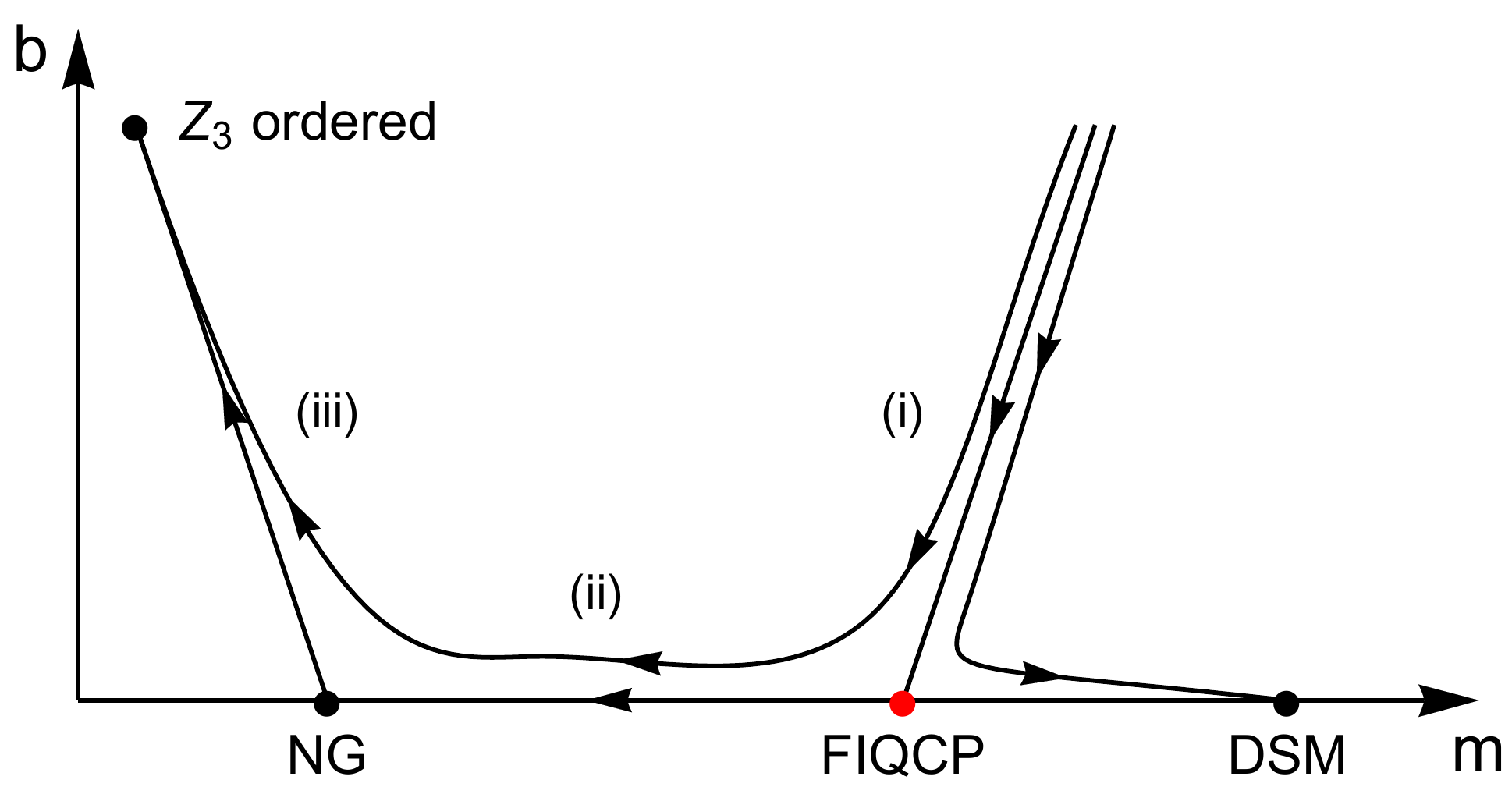}
	\caption{\label{two_scales} A schematic global RG flow diagram. Here the $x$-axis ($y$-axis) represents the tuning parameter $m$ ($Z_3$ anisotropy $b$). The arrows indicate RG flow directions. DSM denotes the Dirac semimetal. FIQCP is the transition point between the semimetal and $Z_3$ ordered phase. NG indicates the Nambu-Goldstone fixed point. Regions (i), (ii), and (iii) denote those regions where scaling behavior is controlled in perturbative RG calculations, e.g., $b^2\ll 1$.}
\end{figure}

\section{Two length scales at the FIQCP}

The existence of dangerously irrelevant fields at a quantum critical point can yield two divergent length scales \cite{oshikawa2000,senthil2004a,carmona2000, senthil2004b, okubo2015, leonard2015, shao2016} and different critical exponents on the two sides of the transition point. A schematic phase diagram is shown in Fig. \ref{two_scales}, where the $x$ axis and $y$ axis represent tuning parameter $m$ and $Z_3$ anisotropy $b$, respectively, and the arrows indicate RG flow directions, i.e., flow to low energy and long wavelength. There are two stable fixed points corresponding to the Dirac semimetal phase and $Z_3$ ordered phase, and the FIQCP is the transition point between them. Moreover, there is another unstable fixed point, the Nambu-Goldstone (NG) fixed point, located on the $b\!=\!0$ axis. If the system is $O(2)$ symmetric, the NG fixed point describes the gapless phase with NG modes resulting from spontaneous breaking of continuous $O(2)$ symmetry in the ordered phase. An effective Lagrangian for the NG phase reads
\bea
	\mathcal{L}= \frac12\sum_{\mu=\tau,x,y} (\partial_\mu \theta)^2,
\eea
where $\theta$ is defined through $\phi\equiv |\phi| e^{i \theta}$. However, since the cubic term $b$ explicitly breaks the $O(2)$ symmetry, $b$ is then relevant at the NG fixed point and drives the system into the $Z_3$ ordered phase where the NG modes are gapped.

When the system moves into the ordered phase and locates slightly away from the critical point with $m\!\lesssim\!m_c$, along the RG flow the system enters region (i), where the scaling behavior in the crossover from region (i) to region (ii) is controlled by the fixed point FIQCP. Here, the characteristic length scale is given by the correlation length of the $Z_3$ order $\xi$, with $\xi \!\sim \!(m_c \!-\!m)^{-\nu}$. As the $Z_3$ anisotropy is dangerously irrelevant at the FIQCP, it is expected that $b^2 \ll 1$ and the system is dominated by the pseudo-NG modes with small masses in region (ii). Flowing to lower energy and longer distance, the system enters region (iii), where the scaling behavior in the crossover from region (ii) to region (iii) is controlled by the NG fixed point instead. Now that the $Z_3$ anisotropy is relevant at the NG fixed point, the pseudo-NG modes are damped, and different domains of $Z_3$ ordering become apparent. Then, another characteristic length scale of the $Z_3$ domain emerges and is given by the diameter of the domains $\xi'$, with $\xi' \!\sim\! (m_c\!-\!m)^{-\nu'}$. These two length scales obey different scaling laws \cite{okubo2015},
\bea
	\frac{\nu'}{\nu}&=&1-\frac{y_{b^2}}{2} \\
&=&1\!+\!\frac{3(\sqrt{N^2+38N+1}\!-\!N\!-\!4)}{10(N+1)},
\eea
where $\nu$ ($\nu'$) is the critical exponent capturing the divergence of characteristic length scales $\xi$ ($\xi'$) approaching criticality from the ordered side, and $y_{b^2}$ is the scaling dimension of $b^2$ at the FIQCP. When the FIQCP occurs (namely, $N\!>\!N_c\!>\!1/2$), one can see that $\nu'\!>\!\nu$. The domain length scale is larger than the correlation length, consistent with our analysis of RG flows above. Note that $\nu'$ exists only in the ordered phase, while the same $\nu$ can be measured on both sides of the transition.

We would like to mention that the emergence of two divergent length scales around the FIQCP differs from the one occurring in the usual bosonic $Z_n$ theory due to the existence of gapless electrons at the FIQCP. In conventional bosonic $Z_n$ theory, the $Z_n$ anisotropy is irrelevant only for $n\! \ge\! 4$ from $4\!-\!\epsilon$ RG calculations \cite{oshikawa2000,hasenbusch2011,okubo2015}. Here, owing to the presence of gapless Dirac fermions at the transition, the $Z_3$ anisotropy is rendered dangerously irrelevant at FIQCP. Moreover, the ratio of $\frac{\nu'}{\nu}$ for the FIQCP depends on the number of flavors of Dirac fermions, as expected.

\section{Conclusions and discussion}

In conclusion, we have presented a RG study of the $Z_3$ Gross-Neveu-Yukawa quantum phase transitions in Dirac systems whose low-energy effective field theory contains cubic terms of order-parameters. Our RG results show that for $d=2$ there exists a finite range of $N$, namely $N\!>\!N_c$, such that the putative first-order phase transitions can be driven to a continuous phase transition due to the fluctuations of gapless Dirac fermions at the transition point. Furthermore, with the help of spacetime SUSY in 2+1D, we obtain an exact lower bound for the critical number of flavors of Dirac fermions to realize a FIQCP, namely $N_c\!>\frac12$. Moreover, the large-scale sign-problem-free QMC simulations \cite{li2015a, wu2016a} indicate that $N_c<2$. We conclude that $\frac12\!<\!N_c\!<\!2$. 

We observe that there is a common feature shared by Landau-forbidden phase transitions (e.g., FIQCPs and DQCPs): the existence of dangerously irrelevant operators. In the language of RG, the Landau criteria can be interpreted by the existence of more than one relevant parameters rendering a first-order phase transition or a multicritical point, rather than a generic quantum critical point. For instance, the $\phi^3$ terms in the case of FIQCPs and the quartic monopole creation operator in the case of DQCPs are seemingly relevant operators (besides the usual relevant mass term of the order parameter). However, these seemingly relevant operators are rendered irrelevant at the critical points by either gapless fermions in the case of FIQCPs or gapless deconfined spinons in the case of DQCPs; as a result, the critical theory exhibits an emergent $O(2)$ symmetry. The presence of dangerously irrelevant operators at the transition point can reduce the emergent symmetry in the symmetry-breaking phases, leading to the emergence of two length scales. We also explored the values of different critical exponents by one-loop RG calculations.

Possible experiments realizing a FIQCP include the quantum phase transitions between semimetallic graphene \cite{novoselov2005, castro2009} and the Kekule-VBS phase \cite{gutierrez2016} by growing the graphene on a certain subtract or by tuning the strain.  There are also experiments on adsorbed monolayer quantum gases, including helium, on the top of graphene-like substrates \cite{bretz1976,reatto2013}, where the quantum gas films can undergo a phase transition to the famous $\sqrt{3}\!\times\!\sqrt{3}$ order, which is a $Z_3$ Gross-Neveu-Yukawa transition.

{\it Acknowledgments}: We would like to thank Yi-Fan Jiang, Zi-Xiang Li, Jiabin Yu and Shi-Xin Zhang for discussions. This work was supported in part by the MOST of China under Grant No. 2016YFA0301001 (H.Y.) and the NSFC under Grant No. 11474175 (S.-K.J. and H.Y).

\appendix
\section{\label{trace}Trace properties for the $\gamma$ matrices}

As defined in the main text, $\gamma^\mu$, $\mu=1,...,d+2$, are the gamma matrices satisfying $\{ \gamma^\mu, \gamma^\nu\}= 2\delta^{\mu\nu}$ and $\gamma^\pm= \frac{1}{2}(\gamma^{d+1}\pm i \gamma^{d+2})$. In the following, we use Greek letters to denote $1,...,d$ and English letters to denote $\pm$. The $\gamma$ matrices defined in the main text have the properties $\{\gamma^\mu,\gamma^\nu\}=2\delta^{\mu\nu}$, $\{\gamma^i,\gamma^j\}=2g^{ij}$, $\{\gamma^i,\gamma^\mu\}=0$, where $g^{ij}=\frac{1}{2}(1-\delta^{ij})$. The traces of the $\gamma$ matrices are given by
\bea
\text{Tr}[\gamma^i\gamma^\mu\gamma^j\gamma^\nu] &=& -4N g^{ij}\delta^{\mu\nu} \nn \\
\text{Tr}[\gamma^\mu \gamma^\nu\gamma^\rho \gamma^\sigma] &=& 4N[\delta^{\mu\nu}\delta^{\rho\sigma}-\delta^{\mu\rho}\delta^{\nu\sigma}+\delta^{\mu\sigma}\delta^{\nu\rho}] \nn \\
\text{Tr}[\gamma^i\gamma^\mu\gamma^j\gamma^\nu\gamma^k\gamma^\rho] &=& 0 \nn \\
  \text{Tr}[\gamma^i\gamma^\mu\gamma^j\gamma^\nu\gamma^k\gamma^\rho\gamma^l\gamma^\sigma] &=&
4N[\delta^{\mu\nu}\delta^{\rho\sigma}-\delta^{\mu\rho}\delta^{\nu\sigma}+\delta^{\mu\sigma}\delta^{\nu\rho}] \nn\\
&& \times [g^{ij}g^{kl}-g^{ik}g^{jl}+g^{il}g^{jk}] \nn
\eea
where we only consider 4$\times$4 $\gamma$ matrices since we are interested in $d=2,3$, and assume the flavor of Dirac fermion to be $N$.

\section{\label{mass_rg}Renormalization of the boson mass term in Dirac semimetals}

The Feynman diagrams contributing to boson mass renormalization are shown in Fig. \ref{fig4} in the main text. A straightforward evaluation leads to
\bea
	\Gamma_{|\phi|^2}&=&-18b^2 \int \frac{d^dk}{(2\pi)^d} \frac{1}{(k^2+r)^2}+ 4u  \int \frac{d^dk}{(2\pi)^d} \frac{1}{k^2+r},\nn \\
&=&\left(- \frac{9\pi\tilde{b}^2}{(1+\tilde{r})^{3/2}} + \frac{4\pi\tilde{u}}{\sqrt{1+\tilde{r}}} \right) \Lambda^2 l ,
\eea
where the various dimensionless coupling constants are defined in the main text. Then, the renormalization-group (RG) equation for boson mass is given by
\bea
\frac{d\tilde{r}}{dl}&=& 2\tilde{r}-\frac{N\pi\tilde{g}^2 \tilde{r}}{2} -\frac{9\pi\tilde{b}^2\tilde{r}}{4(1+\tilde{r})^{5/2}} \nn \\
 &&~~~~~~~~~~~~~~~~~~~~~~  -\frac{9\pi\tilde{b}^2}{(1+\tilde{r})^{3/2}} + \frac{4\pi\tilde{u}}{\sqrt{1+\tilde{r}}}.
\eea
As a first approximation, the solution at the fermion-induced quantum critical point is obtained by substituting the fixed point coupling constants in the above RG equation, and the result yields $r\approx -\frac{8}{N}$.

\end{document}